# Matrix-assisted fabrication and exotic charge mobility of (Li,Fe)OHFeSe superconductor films

Y. L. Huang [1,2,+], Z. P. Feng [1,2,+], J. Yuan [1,3,+], W. Hu [1,2], J. Li [4], S. L. Ni [1,2], S. B. Liu [1,2], Y. Y. Mao [1,2], H. X. Zhou [1], H. B. Wang [4], F. Zhou [1,2], G.M. Zhang [5], K. Jin[1,2,3*], X. L. Dong[1,2,3*], and Z. X. Zhao[1,2,3*]

[1] Beijing National Laboratory for Condensed Matter Physics and Institute of Physics, Chinese Academy of Sciences, Beijing 100190, China

[2] University of Chinese Academy of Sciences, Beijing 100049, China

[3] Key Laboratory for Vacuum Physics, University of Chinese Academy of Sciences, Beijing 100049, China

[4] Research Institute of Superconductor Electronics, Nanjing University, Nanjing 210093, China

[5] State Key Laboratory of Low Dimensional Quantum Physics and Department of Physics, Tsinghua University, Beijing 100084, China

[+] These authors contributed equally
[*] E-mail: dong@iphy.ac.cn; kuijin@iphy.ac.cn; zhxzhao@iphy.ac.cn

**Abstract**

Superconducting $(Li_{1-x}Fe_x)OHFe_{1-y}Se$ films are attractive for both the basic research and practical application. However, the conventional vapor deposition techniques are not applicable in synthesizing the films of such a complex system. So no intrinsic charge transport measurements on the films are available so far to reveal the nature of charge carriers, which is fundamental to understanding the iron-based superconductivity mechanism. Herein we report a soft chemical film technique (matrix-assisted hydrothermal epitaxial growth), by which we have succeeded in growing a series of $(Li_{1-x}Fe_x)OHFe_{1-y}Se$ films covering the whole superconducting regime, with the superconducting transition temperature ($T_c$) from 4 K up to 42 K. This film technique opens up a new way for fabricating other complex functional materials as well. Furthermore, our systematic transport investigation on the film samples indicates that both the electron and hole carriers contribute to the charge transport, with the scattering rates deviating from the Fermi liquid. We find that the superconductivity occurs upon the electron and hole mobility becoming divergent. And in the high $T_c$ samples, the electron carriers are found much more mobile than the holes, a feature distinct from the low $T_c$ samples. Hence, our transport results provide key insights into the underlying physics for iron-based high-$T_c$ superconductivity.

Iron-based superconductor films are emerging as a hot topic with regard to the experimental studies on the superconductivity mechanism and possible applications.[1-12] In particular, it is intriguing that a possible $T_c$ as high as above 65 K has been reported for a monolayer film of binary iron selenide FeSe,[13-18] in contrast to a rather low $T_c$ (~10 K) commonly in multilayer films of FeSe. [5,9,11,19] However, probing into some intrinsic electronic behaviors as charge transport in the monolayer FeSe film is hampered by its high air sensitivity and interface effect. These drawbacks cast a shadow over a comprehensive understanding of this particular kind of material and its application as well. Nevertheless, the recently discovered intercalated iron selenide, $(Li_{1-x}Fe_x)OHFe_{1-y}Se$ (FeSe-11111),[20] has turned out to be the best substitute. It shows a Fermi topology similar to the monolayer FeSe [21,22] and a high $T_c$ over 40 K, even above 50 K under a 12.5 GPa pressure.[23] Compared to the prototypal and other intercalated iron selenides, FeSe-11111 manifests itself as a clean system ideal for investigating the intrinsic electronic properties. It is free from the complications of phase separation, structure transition and interface effects. Moreover, it has better chemical stabilities in air compared to the monolayer FeSe and in high pressures,[23] providing the ease of the property measurements. However, the film materials of $(Li_{1-x}Fe_x)OHFe_{1-y}Se$ cannot be synthesized by conventional high-temperature growth, due to the hydroxyl ion inherent in the compound. This film synthesis problem is now resolved by the soft chemical technique of matrix-assisted hydrothermal epitaxial growth (MAHEG) that we develop and report here.

First of all, the MAHEG technique is capable of producing high-quality single-crystalline $(Li_{1-x}Fe_x)OHFe_{1-y}Se$ films and manipulating the $T_c$ in a wide range from ~4 K up to ~42 K. This film technique opens up a new way for fabricating other complex functional materials as well. Secondly, the high-quality film samples can be well cleaved and patterned into standard Hall bars of a small thickness (~100 nm) that cannot be achieved on single crystal samples (tens of microns). So, a much higher signal-to-noise ratio, improved by two to three orders of magnitude, can be attained. These favorable sample conditions are vital for extracting the transport information essential for understanding the mechanism of iron-based unconventional superconductivity. Actually, we find that both the electron and hole carriers contribute to the electrical transport and their mobilities become diverging as approaching the superconducting transition temperature. Particularly in the high $T_c$ samples, the electron carriers are found much more mobile than the holes. Thirdly, the high-$T_c$ FeSe-11111 film by MAHEG has shown a large critical current density[12] over 0.5 MA/cm$^2$ and high upper critical fields,[12,24,25] thus promising for practical applications in high fields and high-performance electronic devices.

Figure 1a is a schematic picture of the MAHEG process for growing $(Li_{1-x}Fe_x)OHFe_{1-y}Se$ films. Similar to our previous method of the hydrothermal ion-exchange for the optimal superconducting $(Li_{0.84}Fe_{0.16})OHFe_{0.98}Se$ single crystal,[24] *insulating* $K_{0.8}Fe_{1.6}Se_2$ single crystals were used as a matrix. The

K$_{0.8}$Fe$_{1.6}$Se$_2$ precursor, containing FeSe$_4$-tetrahedra layers similar to the target compound, facilitates the hydrothermal epitaxial growth on a substrate of LaAlO$_3$ (LAO) single crystal. The composition, $x$ and $y$, for the superconducting films of different $T_c$'s are estimated to be around 0.2 and less than 0.05, respectively, but cannot be definitely determined by routine probes due to the limited sample mass of each batch of MAHEG (see *Supporting Information*). Nevertheless, the lattice constant $c$ can be accurately determined as a control parameter for the $T_c$ of FeSe-11111.[26] More experimental details can be found in *Supporting Information*, including by other matrix and substrate crystals.

Five typical FeSe-11111 film samples denoted as SC42, SC35, SC30, SC20 and SC04, respectively, were characterized by X-ray diffraction (XRD). The formation of (Li$_{1-x}$Fe$_x$)OHFe$_{1-y}$Se phase, characteristic of a primitive lattice, and a single preferred (001) orientation are confirmed by the measurements of $\theta$-$2\theta$ scan (*Supporting Information* Figure S1). The XRD patterns characteristic of (00$l$) reflections in absence of integral systematic extinctions are in agreement with that of the FeSe-11111 single crystal.[24] But a position shift of the (00$l$) peaks is clearly visible from the zoom-in (005) patterns (Figure 1b), indicative of a gradual $c$-axis lattice expansion from sample SC04 to SC42. The in-plane crystal mosaic of the films is small, in the range of 0.08° to 0.22° (except sample SC04 of the lowest $T_c$) in terms of the full width at half maximum (FWHM) of the rocking curves (Figure 1c). To our knowledge, these FWHM values are so far the best among various iron-based superconducting films and single crystals. The $\varphi$-scan of (101) plane for each sample consists of four successive peaks spaced out by 90° (Figure 1d). It agrees with the $C_4$ symmetry of FeSe-11111, thus demonstrating an excellent out-of-plane orientation and epitaxy. The inset of Figure 1c shows a shiny, mirror-like surface morphology of a cleaved film. These results clearly indicate that the films are of high crystalline quality.

The superconductivity of the films was characterized by in-plane electrical resistivity $\rho_{xx}$ (Figure 2a) and verified by magnetic susceptibility $\chi$ (Figure 2b) measurements. The superconducting transition temperature $T_c$ at zero resistivity is determined as 42.2 K for SC42, 35 K for SC35, 30 K for SC30, 20 K for SC20 and 4 K for SC04. As seen from Figure 2c, the $T_c$ increases with the lattice expansion along $c$-axis, or in other word, with the expansion in the interlayer spacing. Such a positive correlation between the $T_c$ and the interlayer separation was first reported for the powder samples of FeSe-11111.[26] It seems common to a variety of iron selenide superconductors[24] even the pressurized (<5 GPa) FeSe-11111 as well.[23]

Benefiting from the high-quality FeSe-11111 film samples, we are able to attain the electrical transport quantities essential for disentangling the characters of the normal-state charge carriers. The Hall resistivity of each sample is proportional to the magnetic field $B$, i.e. $\rho_{xy}(B) \propto B$ (Figures 3a and S2). Thus, the slope explicitly determines the Hall coefficient $R_H = \rho_{xy}(B)/B$. Figure 3b shows the temperature dependences of $R_H$ for the five samples, with the $R_H$ values all negative at measuring temperatures. The proportional relationship of $\rho_{xy}(B) \propto B$ commonly occurs in electronic systems of either one type or nearly compensated two types of charge carriers. It is obvious that the one-carrier picture *cannot* account for the non-monotonic behavior of $R_H$ vs. $T$ (Figure 3b). In the two-carrier model, the Hall resistivity has a simple expression: $\rho_{xy}(B) = (\mu_h-\mu_e) \cdot B/[(\mu_h+\mu_e)(n \cdot e)]$ (*Supporting Information* SI 1). Here, $n$ is the carrier number, $e$ the electron charge, $\mu_e$ and $\mu_h$ the mobility of electron (e) and hole (h) carriers, respectively. Note that $\rho_{xx}(B) = (1+\mu_h\mu_e B^2)/[(\mu_h+\mu_e)(n \cdot e)]$ (SI 1), thereby the Hall angle can be expressed as $\tan\theta = \rho_{xy}(B)/\rho_{xx}(B) = [(\mu_h-\mu_e)/(1+\mu_h\mu_e B^2)] \cdot B$. So, a relationship of $\tan\theta \propto 1/B$ is expected when $\mu_h\mu_e B^2 \gg 1$, whereas $\mu_h\mu_e B^2 \ll 1$ will yield $\tan\theta \propto B$. For all the film samples, the Hall angle $\tan\theta$ is proportional to $B$ at all measuring temperatures (Figures 4a and S3), which means $\mu_h\mu_e B^2 \ll 1$. In fact, we have performed our measurements in fields no more than 9 T, so that $\mu_h\mu_e B^2 \ll 1$ can be satisfied. The magnetoresistivity can be expressed as MR = $[\rho_{xx}(B)/\rho_{xx}(B=0)]-1 = \Delta\rho_{xx}/\rho_{xx}(B=0) = \mu_h\mu_e B^2$. Actually, a proportional MR $\propto B^2$ holds for all our samples in the normal state up to 80 K (Figures 4b and S4). Consequently, the mobility values of the electron and hole carriers can be extracted from the slopes of $\tan\theta(B)/B$ ($\cong \mu_h-\mu_e$) and MR/$B^2$ (=$\mu_h\mu_e$).

The temperature dependences of the mobility for the two types of carriers are plotted in Figure 5a. Surprisingly, we find that both the electron and hole mobility, $\mu_e$ and $\mu_h$, tend to diverge as the temperature approaches the onset temperature of the superconductivity ($T_c^{onset}$, the values in Figure 2a caption). Such a divergence, rare in unconventional superconductors, suggests the simultaneous occurrences of the superconducting electron pairing and phase coherence. Note that the charge mobility $\mu = (e \cdot \tau)/m^*$, here $\tau$ and $m^*$ are the scattering relaxation time and effective mass of an electron (or a hole), respectively. Therefore, with a finite value of $m^*$, such a mobility behavior reflects a diverging scattering relaxation time which signifies the proximity to the superconducting transition. To get further information on the charge scatterings, we choose a function form of $\mu_{fit} = A \cdot (T - T_c^{onset})^{-\alpha}$ to fit the data of $\mu_{e(h)}$ vs. $T$ (Figure 5a), with $\alpha$ and $A$ as free parameters. The fittings yield a reasonable limited range for the exponent: $\sim 0.6 < \alpha < \sim 1$ (*Supporting Information* Table S1). Here we cannot ascertain whether the electron and hole carriers share a common mobility exponent or not (Table S1). Nevertheless, we find that the electron mobility monotonically increases with $T_c$, by contrast the hole mobility tends to be saturated at higher $T_c$'s (Figure 5b). That is, the difference between the electron and hole mobility, $\mu_e - \mu_h$, is small for the lower $T_c$ samples but becomes pronounced for the higher $T_c$ samples. Furthermore, the fitted $\alpha$ values significantly deviate from the Fermi liquid ($\alpha = 2$).

In addition, the curves of Hall coefficient $R_H$ vs. temperature $T$ display a broad dip feature, with a minimum $R_H$ at a characteristic temperature $T^*$ (Figure 3b). We find that the $T_c$ is promoted with increasing $T^*$ (Figures 3b and 3c), which provides a further clue to the interplay of the charge carriers' character and the superconductivity. From $R_H = (\mu_h-\mu_e)/[(\mu_h+\mu_e) \cdot n \cdot e]$, the negative values of the Hall coefficient indicate that the electron mobility is dominant ($\mu_e > \mu_h$). One may expect that the highly mobile electrons and less mobile holes would play their respective roles in forming the high-$T_c$ superconductivity in the films. Likely, the less mobile hole carriers reflect the presence of antiferromagnetic spin fluctuations in FeSe-11111.[24,29,30] Thus, the positive correlation between the $T_c$ and the temperature scale $T^*$ is an important issue worthy of further experimental studies.

Finally, we make further discussions taking account of the recent high-resolution angle-resolved photoemission (ARPES) experiments on high-$T_c$ FeSe-11111 single crystals[21,22] and monolayer FeSe films.[14-16,18] The carrier number (0.08 – 0.1 per Fe) estimated by ARPES for FeSe-11111 is consistent with our data ($10^{20} - 10^{21}$ cm$^{-3}$) estimated by the transport results. However, the ARPES measurements show that only the electron-like Fermi surfaces are present near the Brillouin zone corners (**M** points) but the hole-like bands near the Brillouin zone center ($\Gamma$ point) are 70 - 80 meV below the Fermi level, for both FeSe-11111 and monolayer FeSe. To reconcile the present transport and the ARPES results, one may consider that the multibands crossing the Fermi level near the **M** points are strongly anisotropic, so that contribute both the electron-like and hole-like charge signals.[31] However, there is another possibility that the hole carriers, much slower and heavier by the present study, are seemingly localized for the ARPES probe. Hence, our transport findings provide new insights into the normal-state electronic interactions in (Li$_{1-x}$Fe$_x$)OHFe$_{1-y}$Se, from which the superconductivity stems at different $T_c$'s subsequent to the concurrent enhancements of the electron and hole mobility. Particularly, the normal-state charge transport in high-$T_c$ (Li$_{1-x}$Fe$_x$)OHFe$_{1-y}$Se is characteristic of a much higher mobility for the electron than the hole carriers. Our results thus put constraints on a universal microscopic theory for the high-$T_c$ superconductivity.

**Acknowledgements**
K.J. would like to thank Profs. T. Xiang and Y.F. Yang for helpful discussions. This work was supported by the National Key Research and Development Program of China (Grant Nos. 2017YFA0303003, 2016YFA0300300 and 2015CB921000), the National Natural Science Foundation of China (Grant Nos. 11574370, 11234006, 11474338, 11674374 and 61501220), the Strategic Priority Research Program and Key Research Program of Frontier Sciences of the Chinese Academy of Sciences (Grant Nos. QYZDY-SSW-SLH001, QYZDY-SSW-SLH008 and XDB07020100), and the Beijing Municipal Science and Technology Project (Grant No. Z161100002116011).

**Author contributions**
X.L.D., K.J. and Z.X.Z. designed the project. Y.L.H. and Z.P.F. grew the films guided by X.L.D. and Z.X.Z.. J.Y. performed the transport measurements with the assistance of Y.L.H., Z.P.F. and J. L.. S.L.N. grew the matrix crystals guided by F.Z.. J.Y., Y.L.H, Z.P.F., J.L. and W.H. characterized the samples. K.J., X.L.D. and F.Z. analyzed and interpreted the transport data with the theoretical support of G.M.Z.. All the authors participated in the discussions.

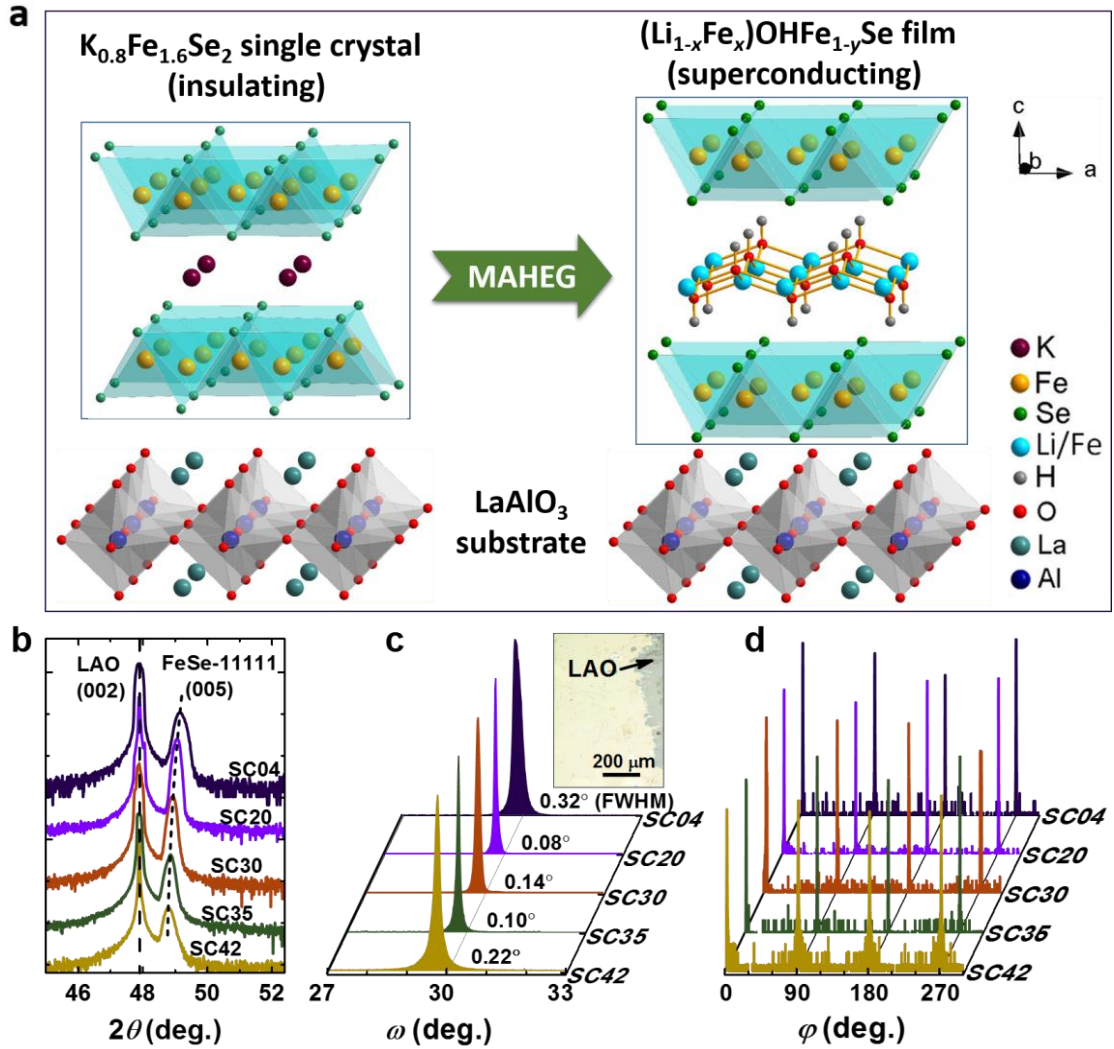

**Figure 1. The film growth and XRD characterizations. a,** Schematic of the matrix-assisted hydrothermal epitaxial growth (MAHEG) on $LaAlO_3$ substrate. Facilitated by the insulating $K_{0.8}Fe_{1.6}Se_2$ matrix crystal (left), the superconducting $(Li_{1-x}Fe_x)OHFe_{1-y}Se$ film (right) is derived under the hydrothermal reaction conditions described in *Supporting Information*. **b,** Zoom-in XRD $\theta$-$2\theta$ scans of the (005) peaks, showing a position shift due to a progressive lattice expansion along $c$-axis from sample SC04 to SC42. **c,** Double-crystal x-ray rocking curves of (006) Bragg reflection and corresponding FWHM values, indicating the small in-plane crystal mosaicity. Inset: photomicrograph of a cleaved film on LAO substrate; the scale bar measures 200 μm. **d,** XRD $\varphi$-scan of (101) plane for each sample consists of four peaks spaced out 90° apart, in agreement with the $C_4$ symmetry. It demonstrates an excellent out-of-plane orientation and epitaxy.

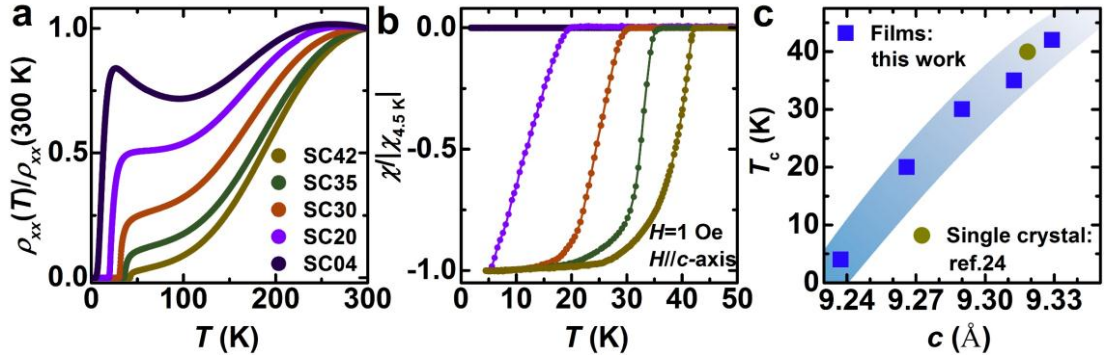

**Figure 2. The superconductivity and its positive correlation with the *c*-axis lattice expansion.** **a**, Temperature dependences of $\rho_{xx}(T)/\rho_{xx}(300\ \text{K})$. The $T_c$ at zero resistivity is determined as 42.2 K for sample SC42, 35 K for SC35, 30 K for SC30, 20 K for SC20, and 4 K for SC04. The $T_c^{onset}$ (onset temperature of the superconductivity) determined from the resistivity is 43.3 K for sample SC42, 40.2 K for SC35, 34.6 K for SC30, 24.7 K for SC20, and 16.9 K for SC04. **b**, Temperature dependences of diamagnetism, measured in zero-field-cooling mode. The signal for sample SC04 of the lowest $T_c = 4$ K is too weak to be detectable. **c**, $T_c$ as a function of the calculated lattice parameter *c* for the films (the blue squares). The data of the FeSe-11111 single crystal[24] (the dark-yellow circular) are included, which conform well to the positive correlation between $T_c$ and *c*.

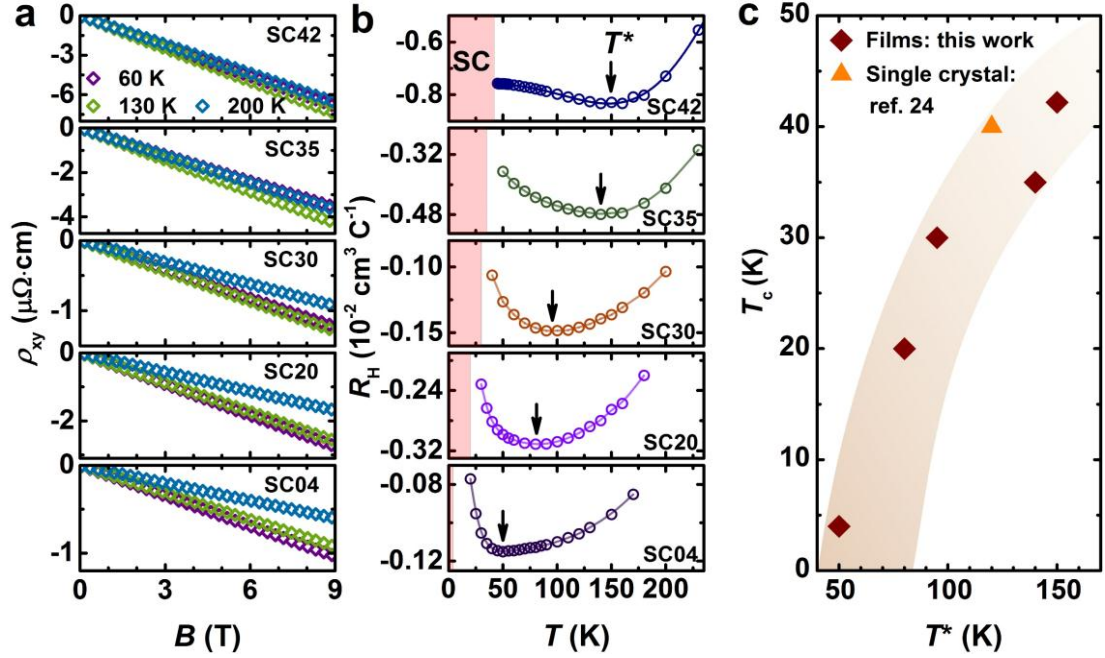

**Figure 3. The Hall resistivity ($\rho_{xy}$) and Hall coefficient ($R_H$) measurements.** **a,** Proportional field dependences of the Hall resistivity $\rho_{xy}(B)$. Here only the data at 60, 130 and 200 K are shown for clarity; the complete data at all measuring temperatures are given in Figure S2. **b,** Temperature dependences of the Hall coefficient $R_H(T)$, showing broad dip feature around characteristic temperatures ($T^*$) marked by the black arrows. The pink shadows indicate the superconducting regions. **c,** Plot of $T_c$ versus $T^*$ for the films (the dark-brown rhombuses), showing the $T_c$ promotion with increasing $T^*$. The corresponding data of the FeSe-11111 single crystal (the orange triangle)[24] are included.

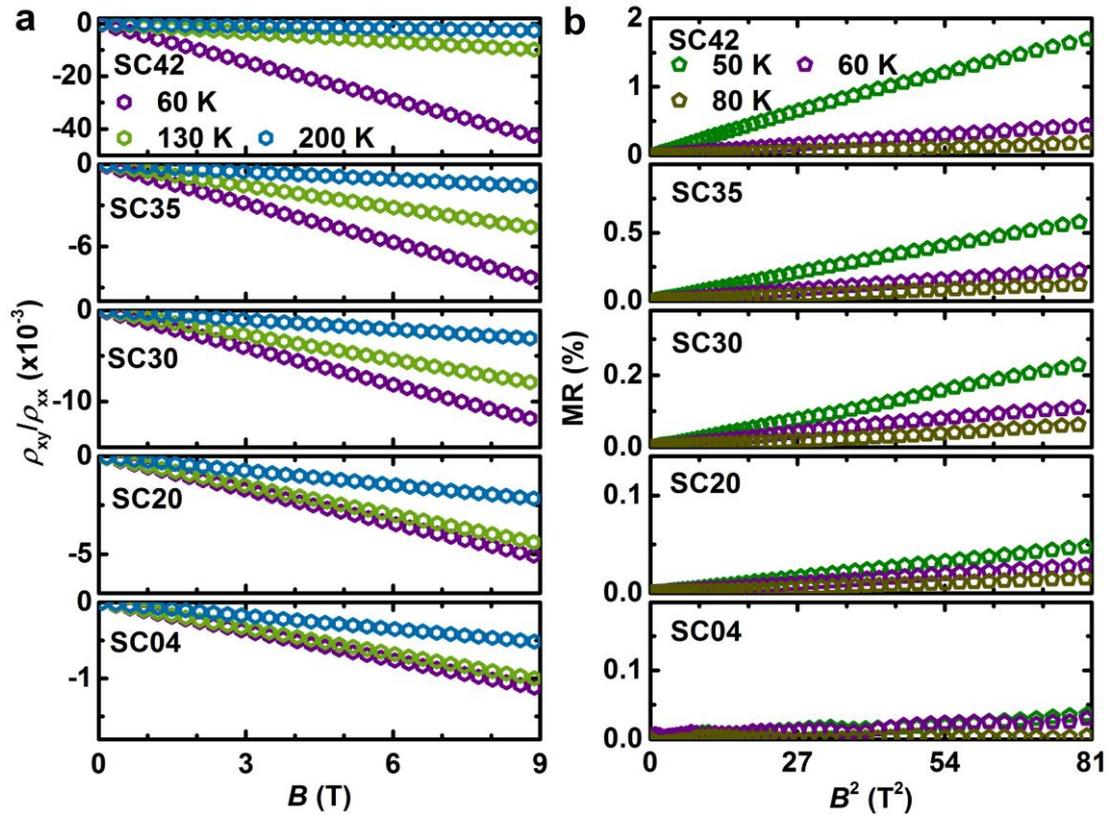

**Figure 4. The Hall angle (tan$\theta$ = $\rho_{xy}/\rho_{xx}$) and magnetoresistivity (MR) measurements.** **a,** Proportional field dependences of the Hall angle at 60, 130 and 200 K. The complete data at all measuring temperatures are given in Figure S3. From the slope of $\rho_{xy}/\rho_{xx} \propto B$, the difference between the hole and electron mobility ($\mu_h-\mu_e$) can be obtained (see text). **b,** Proportional relationship of MR $\propto B^2$ at 50, 60 and 80 K. For the data at all measuring temperatures (below 80 K) see Figure S4. From the slope of MR $\propto B^2$, the quantity of $\mu_h\mu_e$ can be obtained (see text).

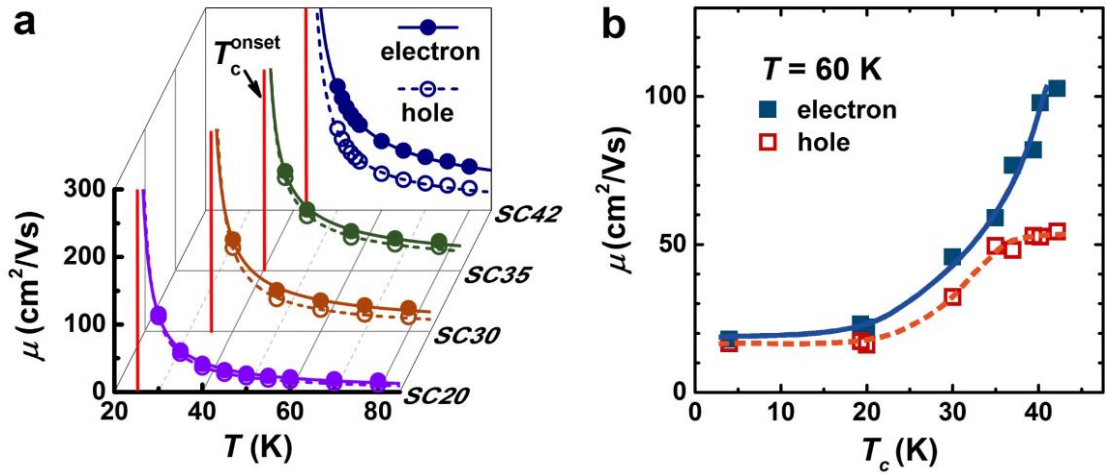

**Figure 5. The evolution of the electron and hole charge mobility.** **a,** Temperature dependences of the electron and hole mobility prior to the temperatures of $T_c^{onset}$, indicated by the red vertical lines. The solid and dashed curves are the fitting results by $\mu_{fit} = A \cdot (T - T_c^{onset})^{-\alpha}$ (see text). Both the electron and hole mobility tend to diverge as approaching $T_c^{onset}$, suggesting the simultaneous occurrences of the superconducting electron pairing and phase coherence. For clarity the data of sample SC04 are not shown. **b,** The electron and hole mobility exhibit distinct $T_c$ dependences. That leads to pronounced mobility differences ($\mu_e - \mu_h$) between the electrons and holes for the higher $T_c$ samples, in contrast to the lower $T_c$ samples. Additional data on film samples besides the five typical ones are included for verification. The solid and dashed curves are guide to the eye.